\documentclass[a4paper]{article}
\usepackage{etex}
\usepackage{a4wide}
\usepackage{graphicx}
\usepackage{epstopdf} 
\usepackage{amsmath,amsbsy,amssymb,amsgen,amsfonts}
\usepackage{color,latexsym}
\usepackage[breaklinks=true,bookmarks=true,colorlinks=true,linkcolor=blue,citecolor=blue,allcolors=blue]{hyperref}
\usepackage[toc,title]{appendix}
\usepackage{natbib}

\usepackage{doi}
\usepackage[section]{placeins} 
\newcommand\caseLa{LC1}         
\newcommand\caseLb{LC2}         
\newcommand\caseTa{TO1}         
\newcommand\Lx{\ensuremath{L_x}}  
\newcommand\Ly{\ensuremath{L_y}}
\newcommand\Lz{\ensuremath{L_z}}
\newcommand\Npart{\ensuremath{N_p}}    
\newcommand\Reb{\ensuremath{Re_b}}   
\newcommand\Ret{\ensuremath{Re_\tau}} 
\newcommand\hfluid{\ensuremath{h_f}}    
\newcommand\hfluidmean{\ensuremath{\langle \overline{h}_f \rangle_{x}}} 
\newcommand\hfluidmeant{\ensuremath{\langle \overline{h}_f\rangle_{xt}}} 
\newcommand\hfluidzmean{\ensuremath{\overline{h}_f }}    
\newcommand\flowrate{\ensuremath{q_f}}    
\newcommand\hbed{\ensuremath{h_b}}    
\newcommand\dratio{\ensuremath{\rho_p/\rho_f}} 
\newcommand\Ga{\ensuremath{Ga}}    
\newcommand\hmean{\ensuremath{H}} 
\newcommand\hbedmean{\ensuremath{\langle \overline{h}_b \rangle_x}}    
\newcommand\hbedzmean{\ensuremath{\overline{h}_b}}    
\newcommand\Dia{\ensuremath{D}}    
\newcommand\Dplus{\ensuremath{D^+}}    
\newcommand\ubulk{\ensuremath{u_b}}    
\newcommand\ufric{\ensuremath{u_\tau}}    

\newcommand\phimeansmooth{\ensuremath{\langle \phi_p \rangle}}
\newcommand\phimeansmoothspanwise{\ensuremath{\langle \phi_p \rangle_z}}
\newcommand\shields{\ensuremath{\Theta}} 

\newcommand\meanlambda{\ensuremath{\lambda_{av}}} 

\newcommand\rmsamplitude{
           \ensuremath{\langle\hbedzmean^\prime\hbedzmean^\prime\rangle_x^{1/2}}}

\def\solidthick{\protect\rule[.5ex]{10.pt}{1.5pt}}

\newcommand{\solidcircle}{$\bullet$}
\newcommand{\solidtriup}{$\blacktriangle$}

\newcommand{\solidsquare}{$\blacksquare$}
\definecolor{mygray}{rgb}{0.7,0.7,0.7}         

\usepackage{sectsty}
\allsectionsfont{\sffamily}
\let\LaTeXmaketitle\maketitle
\renewcommand{\maketitle}{{\sf\LaTeXmaketitle}}

\begin{document}
\epstopdfsetup{suffix=} 
%
\title{Direct numerical simulation of pattern formation in subaqueous sediment}
\author{Aman
  G. Kidanemariam\footnote{\href{mailto:aman.kidanemariam@kit.edu}{aman.kidanemariam@kit.edu}}
  \hspace*{1ex} and
  Markus Uhlmann\footnote{\href{mailto:markus.uhlmann@kit.edu}{markus.uhlmann@kit.edu}}
  \\[1ex]
  $^\ast${\small 
    Institute for Hydromechanics, Karlsruhe Institute of
    Technology}\\ 
  {\small 
    76131 Karlsruhe, Germany
  }
}
\date{} 
\maketitle
%
\begin{abstract}
We present results of direct numerical simulation of incompressible
fluid flow over a thick bed of mobile, spherically-shaped particles. 
The algorithm is based upon the immersed boundary technique for
fluid-solid coupling and uses a soft-sphere model for the solid-solid
contact. 
Two parameter points in the laminar flow regime are chosen, leading to
the emergence of sediment patterns classified as `small dunes', while
one case under turbulent flow conditions leads to `vortex dunes' with
significant flow separation on the lee side. 
Wavelength, amplitude and propagation speed of the patterns extracted
from the spanwise-averaged fluid-bed interface are found to be
consistent with available experimental data.
The particle transport rates are well represented by
available empirical models for flow over a plane sediment bed in
both the laminar and the turbulent regimes. 
\end{abstract}
%
%
%
\section{Introduction}
\label{sec:introduction}
The process of erosion of particles from an initially flat subaqueous
sediment layer and their deposition at certain preferential locations
leads, under certain circumstances, to the amplification of small
perturbations and gives rise to wave-like bed shapes which are
commonly described as ripples or dunes. These sedimentary patterns are
commonly observed in river and marine flows, as well as in various
technical applications involving shear flow over a bed of mobile sediment
particles. From an engineering point of view it is highly desirable to
be able to predict the occurrence and the nature of this phenomenon,
since the bed-form significantly influences flow characteristics such
as resistance, mixing properties and sediment transport.
%

Most of the previous theoretical work on the formation of sediment
patterns is based upon the notion that a flat bed is unstable with
respect to perturbations of sinusoidal shape. It was
\citet{Kennedy1963} who first studied this instability problem by
considering a potential flow solution,
 and over the years the concept
has been applied by a number of researchers for a variety of flow
conditions, in the laminar \citep{Charru2002,Charru2006d} and
turbulent regime
\citep{Richards1980,Sumer1984,Colombini2004,Colombini2011}.  
Invoking a disparity in time scales between the flow and the bed shape 
modification, most of the approaches have considered the bed shape as
fixed for the purpose of the analysis.  The hydrodynamic stability
problem is then complemented by an expression for the particle flux as
a function of the local bed shear stress at a given transversal
section of the flow.  

It is now generally accepted that the mechanism
which destabilizes a 
flat sediment bed is the phase-lag between the
perturbation in bed height and the bottom shear stress 
as a consequence of fluid inertia. 
A balance between this destabilizing mechanism and
other stabilizing effects such as gravity \citep{Engelund1982} or
phase-lag between bottom shear stress and the particle flow rate
\citep{Charru2006c} is believed to result in instability of the bed at
a certain preferred wavelength.  
Linear stability analysis is 
often applied to the problem in order to predict 
the most unstable wavelength; compared to
experimental observations, however, predictions 
resulting from this 
approach 
can be broadly described as unsatisfactory, sometimes
predicting pattern wavelengths which are off by an order of magnitude
\citep{Raudkivi1997,Langlois2007a,Coleman2009,Ouriemi2009b}.

Most available experimental studies report wavelengths of the
developed bed-forms 
after they have undergone a coarsening process 
(the temporal evolution of the initial patterns to their `steady-state'
form), or possibly after they have coalesced with other
bed-forms. There are several experimental studies
which report on the initial wavelength and its development
\citep[][]{Coleman1994,Betat2002a,Coleman2003,Langlois2007a,Ouriemi2009b}. 
However, the reported data is widely dispersed.  
Today it is still a challenge to capture the 
three-dimensional nature of the individual 
particle and fluid motion within the bed layer
in a laboratory experiment, 
which in turn has hindered the assessment of the various theoretical
approaches. 

In the present work 
we numerically investigate the development of subaqueous
patterns in a statistically uni-directional channel flow configuration
both in the laminar and turbulent regimes.  
A sufficiently large number of freely-moving spherical particles are 
represented such that they form a realistic sediment bed in the
simulation.  
To our knowledge, no attempt to numerically simulate the evolution
of a bed of mobile sediment particles (leading to pattern formation)
by means of direct numerical simulation (DNS), which resolves all the
relevant length and time scales of the turbulent flow as well as the
individual sediment particles, has been reported to the present
date.  The present study focuses on aspects related to the initial bed
instabilities and 
their subsequent short-time development. 
%
%
\begin{figure}
  \begin{minipage}{.27\linewidth}
    \centerline{$(a)$}
    \includegraphics[width=\linewidth,clip=true,
    viewport=105 385 420 700]
    {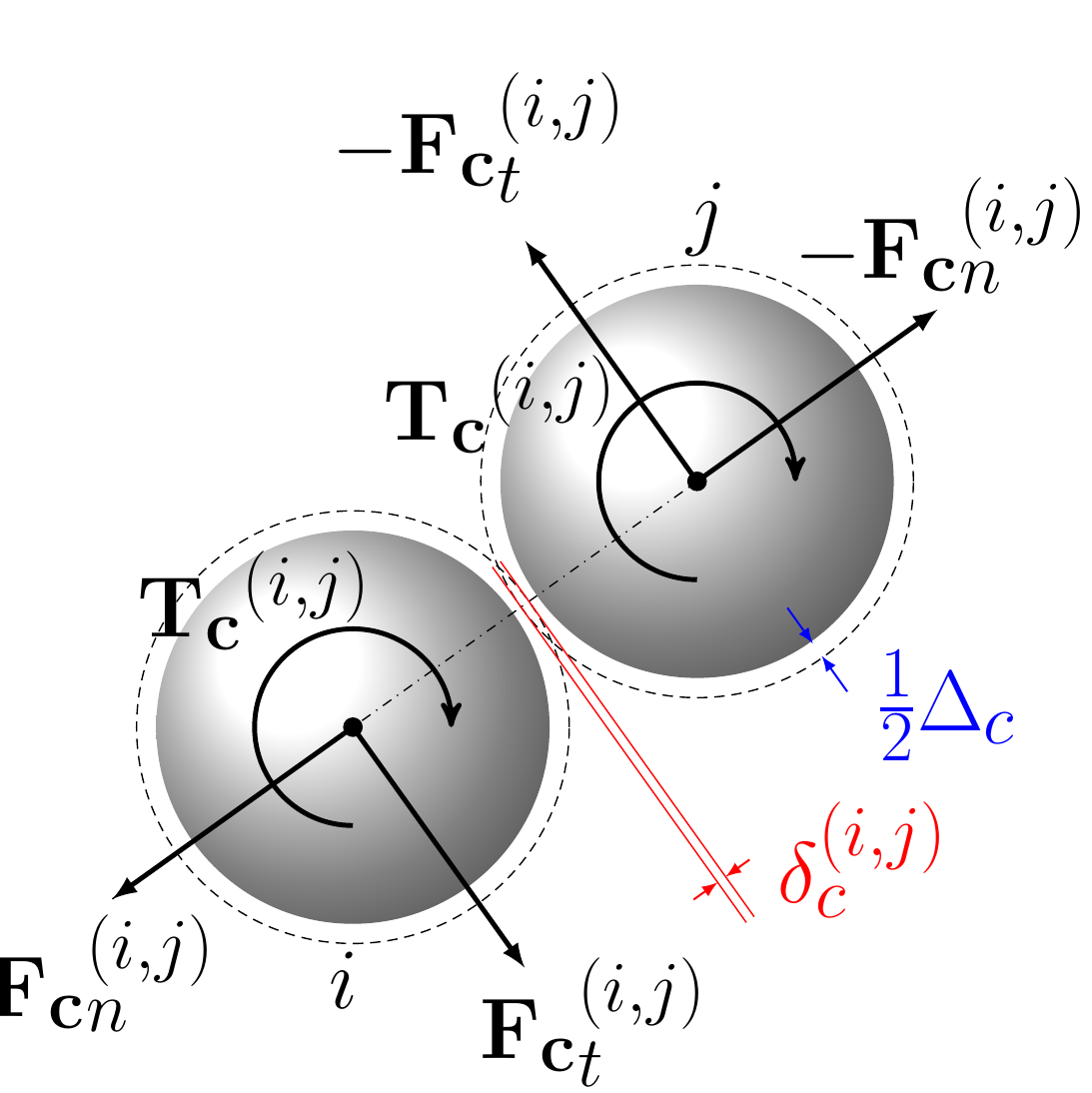}
  \end{minipage}
  \hfill
  \begin{minipage}{0.33\linewidth}
    \centerline{$(b)$}
    \includegraphics[width=\linewidth,clip=true,
    viewport=165 330 520 630]
    {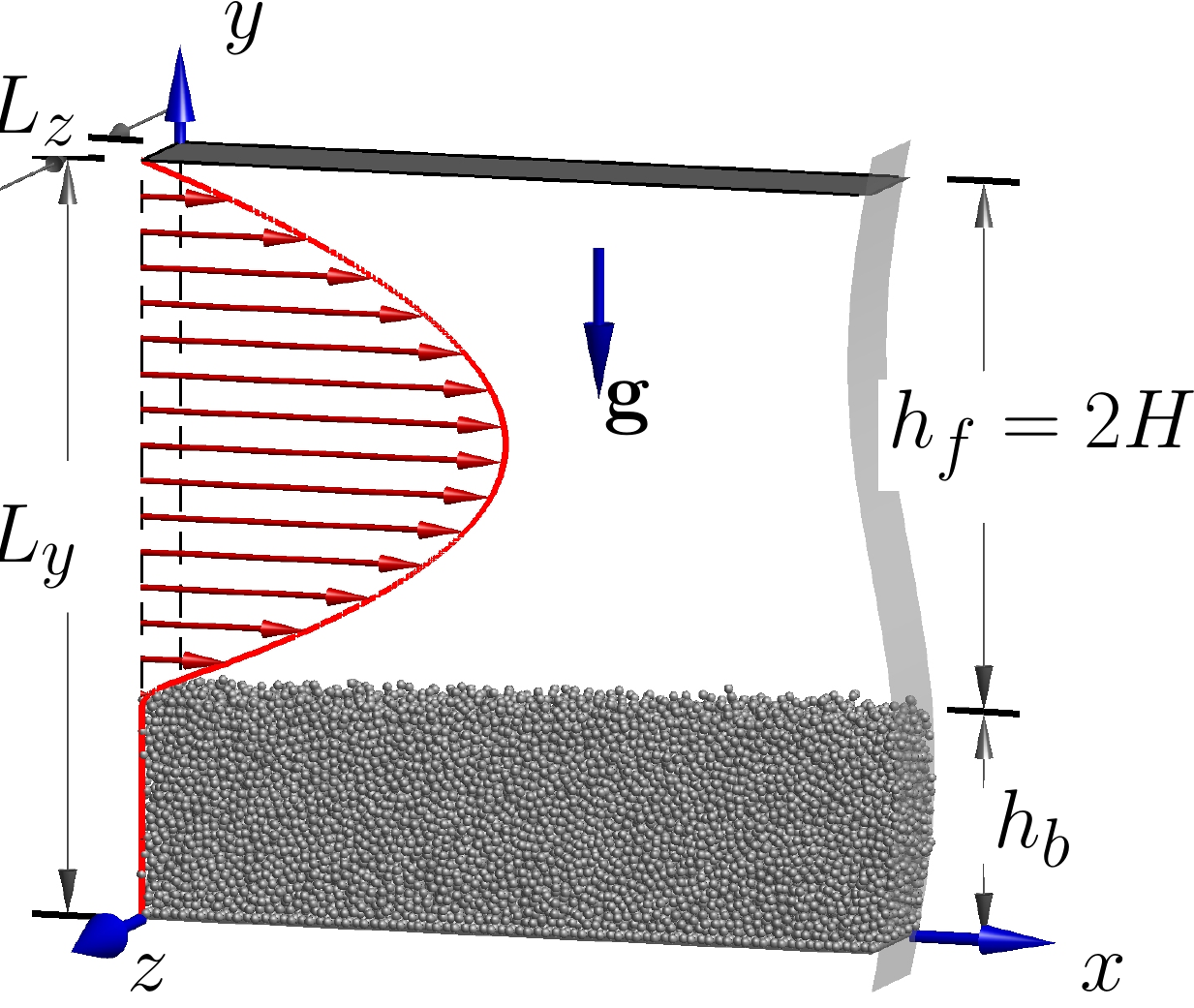}
  \end{minipage}
  \begin{minipage}{0.38\linewidth}
    \centerline{$(c)$}
    \includegraphics[width=\linewidth,clip=true,
    viewport=120 300 560 640]
    {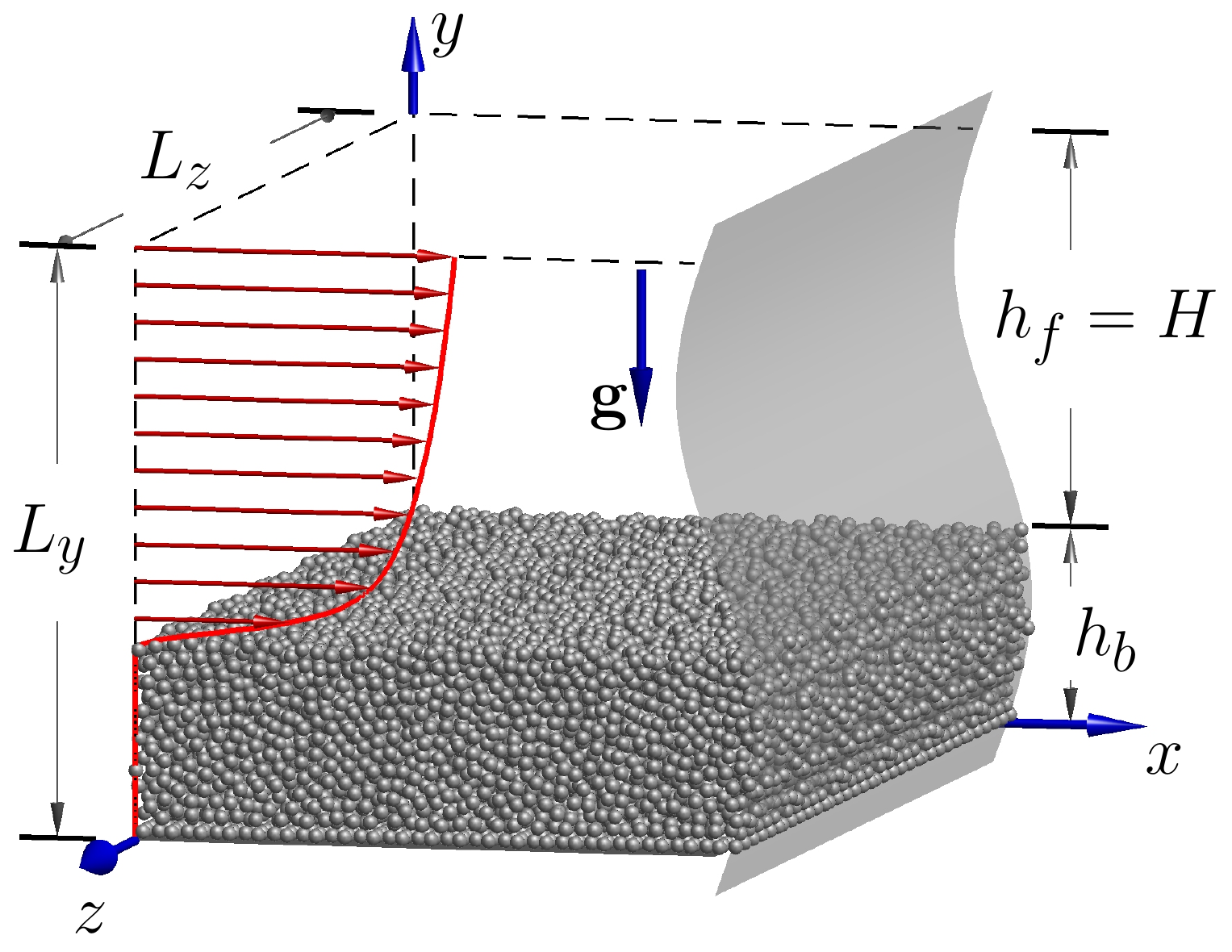}
  \end{minipage}
  \caption{%
    (\textit{a}) 
    Schematic showing the collision force $\mathbf{F_c}^{(i,j)}$ and torque
    $\mathbf{T_c}^{(i,j)}$ acting on particle with index $i$ during
    contact with particle $j$. The subscripts $n$ and $t$
    indicate the normal and tangential components, respectively; 
    $\Delta_c$ denotes the force range, and $\delta_c^{(i,j)}$ is the
    overlap length. 
    (\textit{b}) 
    Computational domain and coordinate definition in cases \caseLa\
    and \caseLb.  
    (\textit{c}) 
    The same for case \caseTa.
    The computational domains are periodic along the $x$- and
    $z$-directions with periods $L_x$, $L_z$, respectively. 
  }
  \label{fig:schematic_diagram}
\end{figure}
%
\section{Computational setup}
\label{sec:computational-setup}
\subsection{Numerical method}
\label{subsec:numerical-method}
The numerical treatment of the fluid-solid system is based upon the
immersed boundary technique. 
The incompressible Navier-Stokes equations are solved with a
second-order finite-difference method throughout the
entire computational domain $\Omega$ (comprising the fluid domain
$\Omega_{\mathrm{f}}$ and the domain occupied by the suspended
particles $\Omega_{\mathrm{s}}$), adding a localized force term which
serves to impose the no-slip condition at the fluid-solid interface. 
The particle motion is obtained via integration of the Newton
equation for rigid body motion, driven by the hydrodynamic force 
(and torque) as well as gravity and the force (torque)
resulting from inter-particle contact. 
Further information on the extensive validation
of the direct numerical simulation (DNS) code on a whole range of 
benchmark problems can be found in
\cite{Uhlmann2005a}, \cite{uhlmann:13a-nodoi} and further
references therein. 
The code has been previously employed for the simulation of various
particulate flow configurations
\citep[][]{Uhlmann2008,Chan-braun2011,Garcia-villalba2012,
Kidanemariam2013}. 

In the present case, direct particle-particle contact contributes
significantly to the dynamics of the system. 
In order to realistically simulate the collision process between the 
immersed particles, a discrete element model (DEM) based
on the soft-sphere approach is coupled to the two-phase flow solver.
The DEM used in the present work employs a linear
mass-spring-damper system to model the collision forces, which are
computed independently for each colliding particle pair. 
Any 
two particles are defined as `being in contact' 
when the smallest distance between their 
surfaces, $\Delta$,
becomes smaller 
than a force range $\Delta_c$ 
as illustrated in figure~\ref{fig:schematic_diagram}(\textit{a}). 
%
The collision force is computed from the sum of an elastic normal
component, a normal damping component and a tangential frictional
component. 
The elastic part of the normal force component 
is a linear function of the penetration length  
$\delta_c \equiv \Delta_c-\Delta$, with a stiffness constant $k_n$. 
The normal damping force is a linear function of the normal component 
of the relative velocity between the particles at the 
contact point with a constant coefficient $c_n$. 
The tangential frictional force (the magnitude of which is limited 
by the Coulomb friction limit with a friction coefficient $\mu_c$) is
a linear function of the tangential 
relative velocity at the contact point, again formulated with a
constant coefficient denoted as $c_t$. 
A detailed description of the collision model
and extensive validation 
tests with respect to available experimental data of a single particle
colliding with a wall in a viscous fluid and 
in the case of bedload transport under laminar shear flow 
has been recently provided by 
\citet{kidanemariam:14a}. 
%


%
\begin{table}
  \begin{center}
  \begin{tabular}{lccccccc|cccc}
      Case &
      \Reb  & 
      \Ret  &   
      \dratio &
      \Ga & 
      \Dplus &
      $\hmean/\Dia$ &
      $\shields$ &
      $[\Lx \times \Ly \times \Lz]/\Dia$ &
      $\Delta x^+$  &
      $\Dia/\Delta x$ & 
      \Npart \\[3pt]
      \caseLa &
      700 &
      32.31 &
      2.5 &
      2.42 &
      1.20 &
      26.92 &
      0.25 &
      $307.2\times 76.8\times 12.8$ &
      0.12 &
      10 &
      79073 \\
      \caseLb &
      700 &
      32.08 &
      2.5 &
      1.97 &
      1.20 &
      26.71 &
      0.37 &
      $307.2\times 76.8\times 12.8$ &
      0.12 &
      10 &
      79073\\
      \caseTa &
      6022 &
      290.34 &
      2.5 &
      28.37 &
      11.59 &
      25.05 &
      0.17 &
      $307.2\times 38.4\times 76.8$ &
      1.16 &
      10 &
      263412
  \end{tabular}
  \caption{Physical and numerical parameters of the simulations:
    \Reb\ and \Ret\ are the bulk and friction Reynolds numbers, 
    respectively, \dratio\ is the particle-to-fluid density
    ratio, \Ga\ is the Galileo number, \Dia\ is the particle diameter,
    \hmean\ is the equivalent boundary layer thickness, 
    \shields\ the Shields number,
    $L_i$ is the domain size in the $i$th coordinate direction, 
    $\Delta x$ is the uniform mesh width and 
    \Npart\ is the number of particles.
    %
  }
  \label{tab:physical-and-numerical-parameters}
\end{center}
\end{table}
%
The four parameters which describe the collision process in the
framework of this model ($k_n$, $c_n$, $c_t$, $\mu_c$) as well as the
force range $\Delta_c$ need to be prescribed for each simulation.  
From an analytical solution of the linear mass-spring-damper system 
in an idealised configuration
(considering a binary collision of uniformly translating spheres in
vacuum and in the absence of external forces), 
a relation between the normal stiffness 
coefficient $k_n$ and the normal damping coefficient $c_n$ can be
formed by introducing the dry restitution coefficient
$\varepsilon_d$. 
This latter quantity is a material property, defined as the absolute
value of the ratio between the normal components of the relative
velocity post-collision and pre-collision. 
In the present simulations, $\Delta_c$ is set equal to 
one grid spacing $\Delta x$.  
%
%
%
The stiffness parameter $k_n$ has a value equivalent to approximately 
$17000$ times the submerged weight of the particles, divided by
the particle diameter. 
The chosen value ensures that the maximum overlap $\delta_c$ over
all contacting particle pairs is 
within a few percent of
$\Delta_{c}$.  
%
The dry coefficient of restitution is set to $\varepsilon_d=0.3$ 
which together with $k_n$ fixes the value for $c_n$. 
Finally, the tangential damping coefficient $c_t$ was set equal to
$c_n$, and a value of $\mu_{c}=0.4$ was imposed for the Coulomb
friction coefficient.   
This set of parameter values for the contact model was successfully
employed in the simulations of (featureless) bedload transport by
\cite{kidanemariam:14a}. 
In order to account for the large disparity between the
time scales of the particle collision process and those of the
smallest flow scales, 
the Newton equation for particle motion is solved with a
significantly smaller time step than the one used for solving the
Navier-Stokes equations (by a factor of approximately one hundred),
while 
keeping the hydrodynamic contribution to the force and torque acting
upon the particles constant during the intermediate interval.  
%
%
\begin{figure}
  \centering
  \begin{minipage}{2ex}
    \rotatebox{90}
    {\small $Re_b$, $Re_{pipe}$}
  \end{minipage}
  \begin{minipage}{.43\linewidth}
    \includegraphics[width=\linewidth]
    {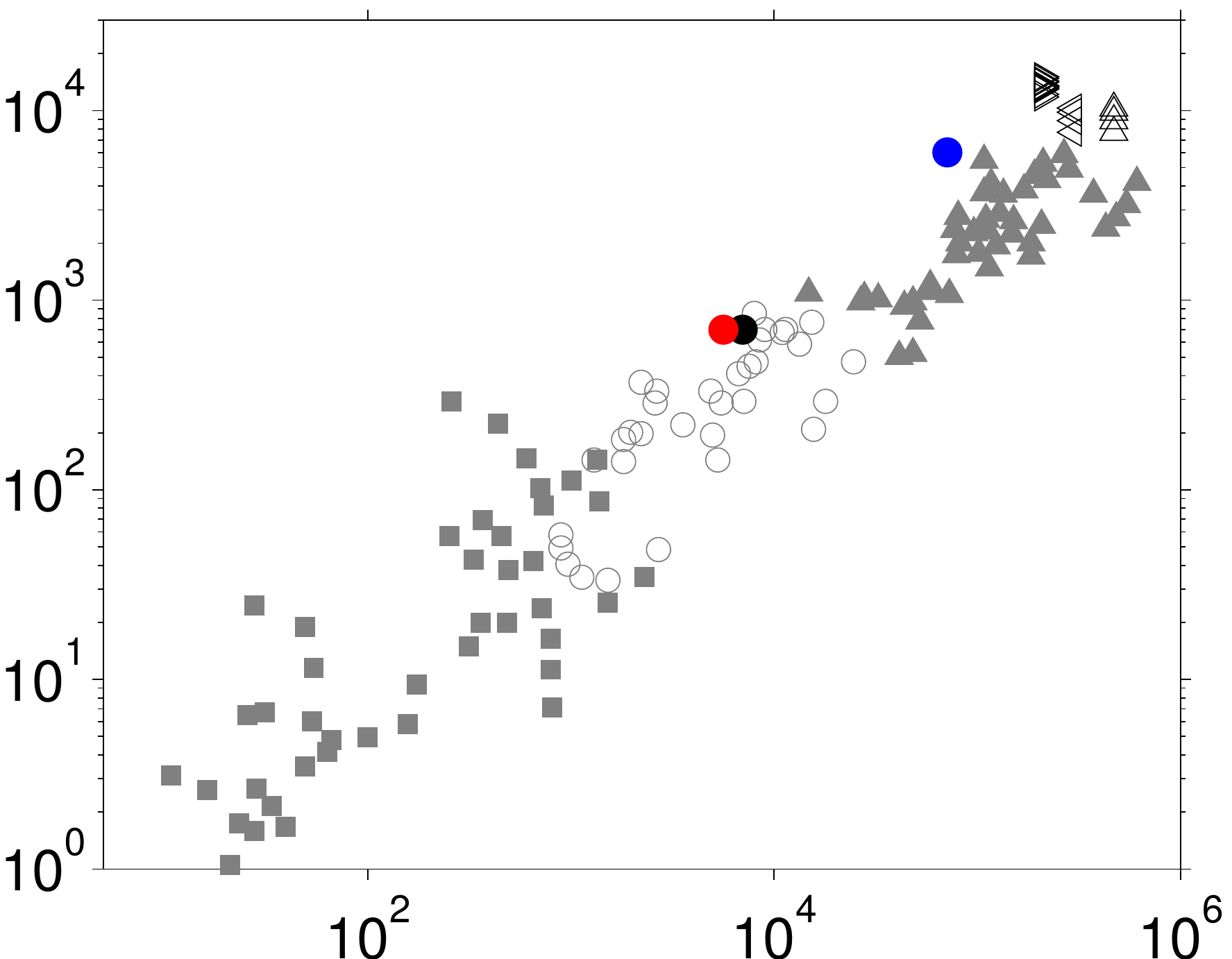}
    \centerline{\small $\Ga ({2\hmean}/{\Dia})^2$}
  \end{minipage}
  \caption{
    %
    Different regimes of sediment bed patterns obtained in the pipe
    flow experiment of \citet{Ouriemi2009b}, shown in the 
    parameter plane ($Re_b$,$\Ga ({2\hmean}/{\Dia})^2$): 
    `flat bed in motion'
    ({\color{mygray}\solidsquare});
    `small dunes' ({\color{mygray}$\circ$});
    `vortex dunes' ({\color{mygray}\solidtriup}).
    %
    %
    For the pipe flow data the Reynolds number
    $Re_{pipe}$ 
    based upon the pipe diameter $d_{pipe}$ 
    and the bulk velocity $q_f/d_{pipe}$ 
    is used. 
    The data points in the turbulent channel flow experiment of 
    \citet{Langlois2007a} are indicated by: 
    $D=100\mu m$ ($\vartriangle$); 
    $D=250\mu m$ ($\vartriangleleft$); 
    $D=500\mu m$ ($\vartriangleright$). 
    The following symbols refer to the present simulations: 
    {\color{black}\solidcircle},~\caseLa;
    {\color{red}\solidcircle},~\caseLb;
    {\color{blue}\solidcircle},~\caseTa.
  }
  \label{fig:parameter-space}
\end{figure}
%
\subsection{Flow configuration and parameter values}
\label{subsec:flow-configuration-and-parameter-values}
The flow considered in this work is horizontal plane channel flow in a
doubly-periodic domain 
as shown in figure~\ref{fig:schematic_diagram}$(b,c)$. 
%
%
%
Three simulations are
performed, henceforth denoted as case \caseLa, case~\caseLb\ 
(both in the laminar flow regime) and case \caseTa\ (in the turbulent 
regime). 
%
In cases \caseLa\ and \caseLb\ the domain is bounded in the vertical
direction by two solid wall planes,  
whereas for reasons of computational cost in case~\caseTa\ 
an open channel is simulated, i.e. 
only the lower boundary plane corresponds to a no-slip
wall, while a free-slip condition is imposed at the upper boundary plane. 
As shown in figure~\ref{fig:schematic_diagram} the Cartesian
coordinates 
$x$, $y$, and $z$ are aligned with the streamwise,
wall-normal and spanwise directions, respectively, 
while gravity acts in the negative $y$-direction. 
The flow is driven by a horizontal pressure gradient 
at constant flow rate \flowrate\ (per unit spanwise length) 
which results in a shear flow of 
height \hfluid\ over a mobile bed of height \hbed; 
spatial averages \hfluidmean\ and \hbedmean\ of both quantities are
defined in \S~\ref{determination-of-the-fluid-bed-interface}; 
temporal averaging over the final period of the simulations is
henceforth indicated by the operator $\langle\cdot\rangle_t$. 
%
%
The bulk Reynolds number of the
flow is defined as $\Reb=2\hmean\ubulk/\nu$, where 
$\ubulk\equiv\flowrate/\hfluidmeant$ is the bulk velocity,  
%
\hmean\ is the equivalent boundary layer thickness 
(i.e.\ $\hmean=\hfluidmeant/2$ in cases \caseLa, \caseLb\ and
$\hmean=\hfluidmeant$ in case \caseTa,
cf.\ figure~\ref{fig:schematic_diagram}$b,c$), 
and $\nu$ is the kinematic viscosity. 
Similarly, the friction Reynolds number is defined as
$\Ret=\ufric\hfluidmeant/\nu$,  
where the friction velocity \ufric\   
is computed by extrapolation of the total
shear stress 
to the 
fully-developed value of the 
%
%
wall-normal location
of the average fluid-bed interface  \hbedmean.
%
Further physical parameters are 
the ratio of particle to fluid density, \dratio, 
the Galileo number 
$\Ga = u_g\Dia/\nu$
(where $u_g=((\dratio -1)|\mathbf{g}|D)^{1/2}$ and \Dia\ the
particle diameter),  
the Shields number 
$\shields= \ufric^2/u_g^2$ 
and the length scale ratio
$\hmean/\Dia$; 
these together with the chosen numerical parameters are shown in 
table \ref{tab:physical-and-numerical-parameters}. 
%
  The present simulations consumed a total of approximately
  5 million core hours on the computing system
  SuperMUC at LRZ M\"unchen. Typical runs of case~\caseTa\ were
  carried out on 576 cores.

Figure~\ref{fig:parameter-space} shows the three parameter points of the
present simulations in the plane spanned by $Re_b$ and 
$\Ga({2\hmean}/{\Dia})^2$) in comparison to the laboratory
experiments of \citet{Ouriemi2009b} and \citet{Langlois2007a}. 
Note that the former experiment was performed in pipe flow, whereas
the latter was in plane channel flow. 
It can be seen that the cases \caseLa\ and \caseLb\ fall into the regime
where the formation of `small dunes' is expected while `vortex dunes'
can be anticipated in case~\caseTa. 

%
%
%
\begin{figure}
        %
  \raisebox{4ex}{
        \begin{minipage}{2.5ex}
          $(a)$\\[3ex]
          {\small $\displaystyle\frac{y}{D}$}
          \\[5ex]
          $(b)$\\[3ex]
          {\small $\displaystyle\frac{y}{D}$}
        \end{minipage}
        }
        \begin{minipage}{.45\linewidth}
          \includegraphics[width=\linewidth]
          {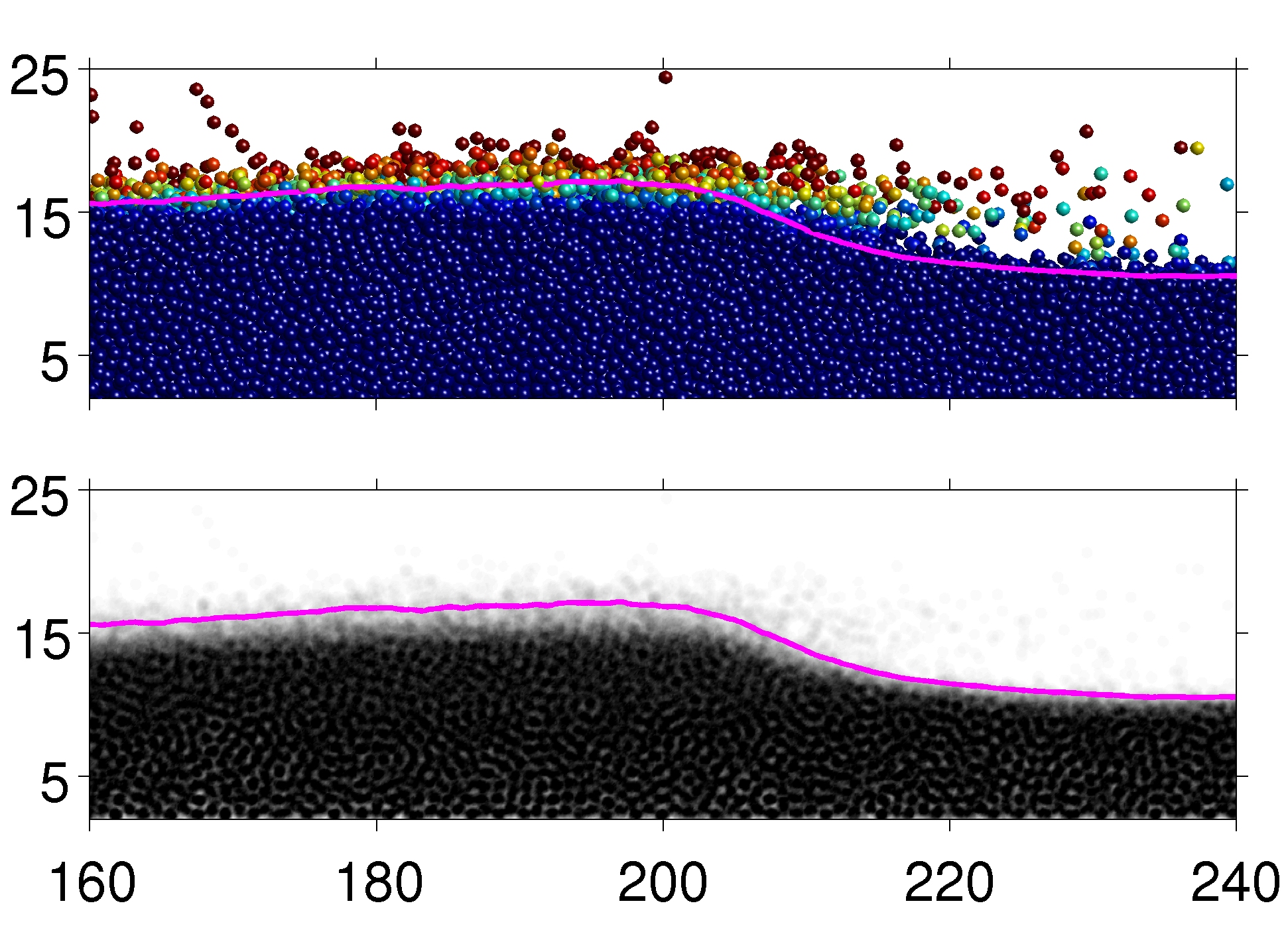}
          \centerline{\small \mbox{$x/D$}}
        \end{minipage}
        \hspace{1ex}
  \raisebox{6ex}{
        \begin{minipage}{2.5ex}
          $(c)$\\[3ex]
          {\small $\displaystyle\frac{y}{D}$}
          \\[7.5ex]
          $(d)$\\[3ex]
          {\small $\displaystyle\frac{y}{D}$}
        \end{minipage}
        }
        \begin{minipage}{.45\linewidth}
          \includegraphics[width=\linewidth]
          {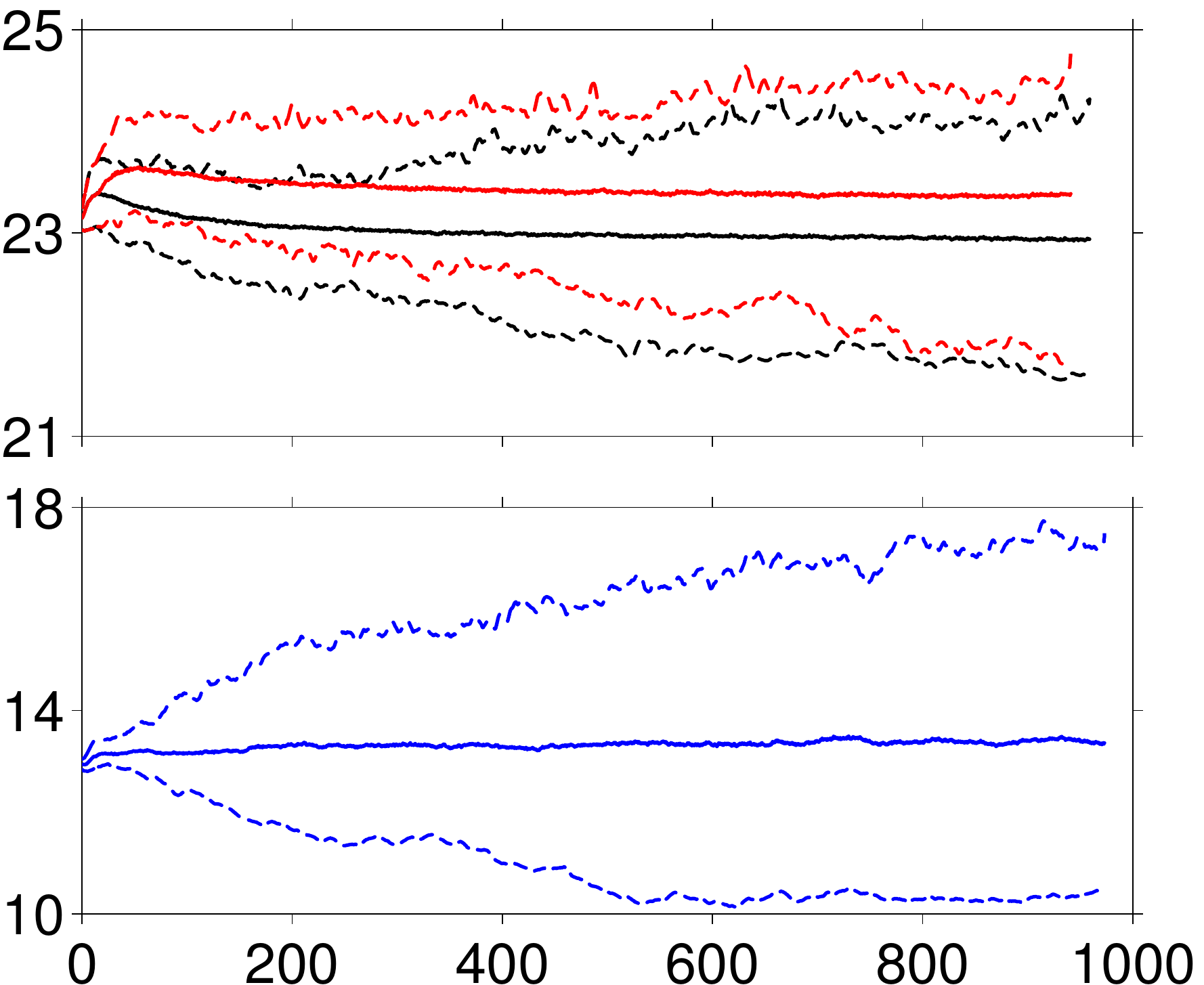}
          \centerline{\small
            $t\,\ubulk/\hmean$}
        \end{minipage}
        \caption{
                 (\textit{a}) Close-up showing 
                 an instantaneous distribution of 
                 particles
                 (colored according to their streamwise velocity)
                 in case \caseTa. 
                 %
                 (\textit{b}) 
                 The spanwise-averaged
                 two-dimensional solid volume fraction 
                 $\phimeansmooth_z(x,y,t)$ corresponding to the snapshot
                 shown in (\textit{a}) displayed in greyscale. 
                 The extracted fluid-bed interface location
                 $\hbedzmean(x,t)$ is shown
                 with a magenta colored line.
                 (\textit{c}) Time evolution of the bed height: 
                 $\hbedmean(t)$ is shown as solid lines, 
                 $\min_x(\hbedzmean)$ and $\max_x(\hbedzmean)$ are
                 shown as dashed lines. 
                 Black color is used for case~\caseLa, 
                 red color for case~\caseLb. 
                 %
                 (\textit{d}) The same quantities as shown in
                 (\textit{c}), but for case \caseTa 
                 %
                 (see also the movies available as supplementary
                 material).
               }
        \label{fig:bed_thickness_evolution}
\end{figure}
%
\subsection{Initiation of the simulations}
\label{subsec:initiation-of-the-simulations}
The simulations were initiated as follows. 
In a first stage the initial sediment bed was generated by means of a
simulation of particles settling (from random initial positions) under
gravity and under solid-solid collisions but disregarding hydrodynamic
forces.  
The result is a pseudo-randomly packed bed of initial bed
thickness $\hbed(t=0)$ above the bottom wall.
%
%
Then the actual fully-coupled fluid-solid simulation is
started with all particles being initially at rest. 
In cases \caseLa\ and \caseLb, the initial fluid velocity field is set
equal to a laminar Poiseuille flow profile with
the desired flowrate in the interval 
$\hbed(t=0)\leq y\leq L_y$ and zero elsewhere. 
After starting the simulation, individual particles are set into
motion due to the action of hydrodynamic force/torque, and erosion
takes place.  
In case \caseTa\ the fluid-solid simulation was first run with
all particles held fixed in order to develop a fully-turbulent field
over the given sediment bed. 
%
After approximately $100$ bulk time units 
the particles were released, and the bed started to evolve away from
its initial macroscopically flat shape, as discussed in the following. 
\begin{figure}
        \centering
        \raisebox{20ex}{$(a)$}
        \begin{minipage}{2ex}
          \rotatebox{90}
          {\small $t\,\ubulk/\hmean$}
        \end{minipage}
        \begin{minipage}{0.25\linewidth}
          \includegraphics[width=\linewidth]
          {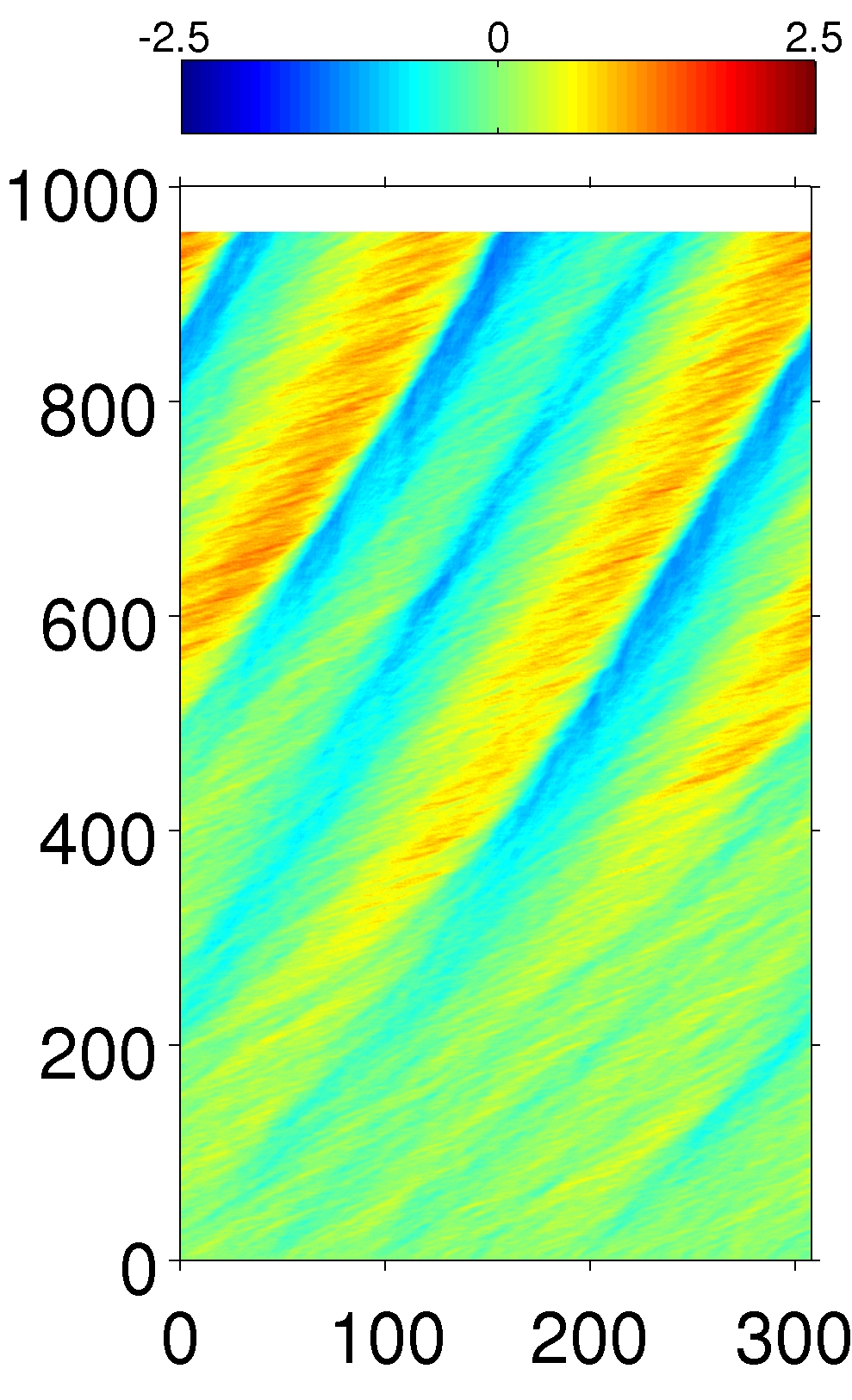}
          \centerline{\small $x/\Dia$ }
        \end{minipage}
        \hspace{1ex}
        \raisebox{20ex}{$(b)$}
        \begin{minipage}{2ex}
          \rotatebox{90}
          {\small $t\, \ubulk/\hmean$}
        \end{minipage}
        \begin{minipage}{0.25\linewidth}
          \includegraphics[width=\linewidth]
          {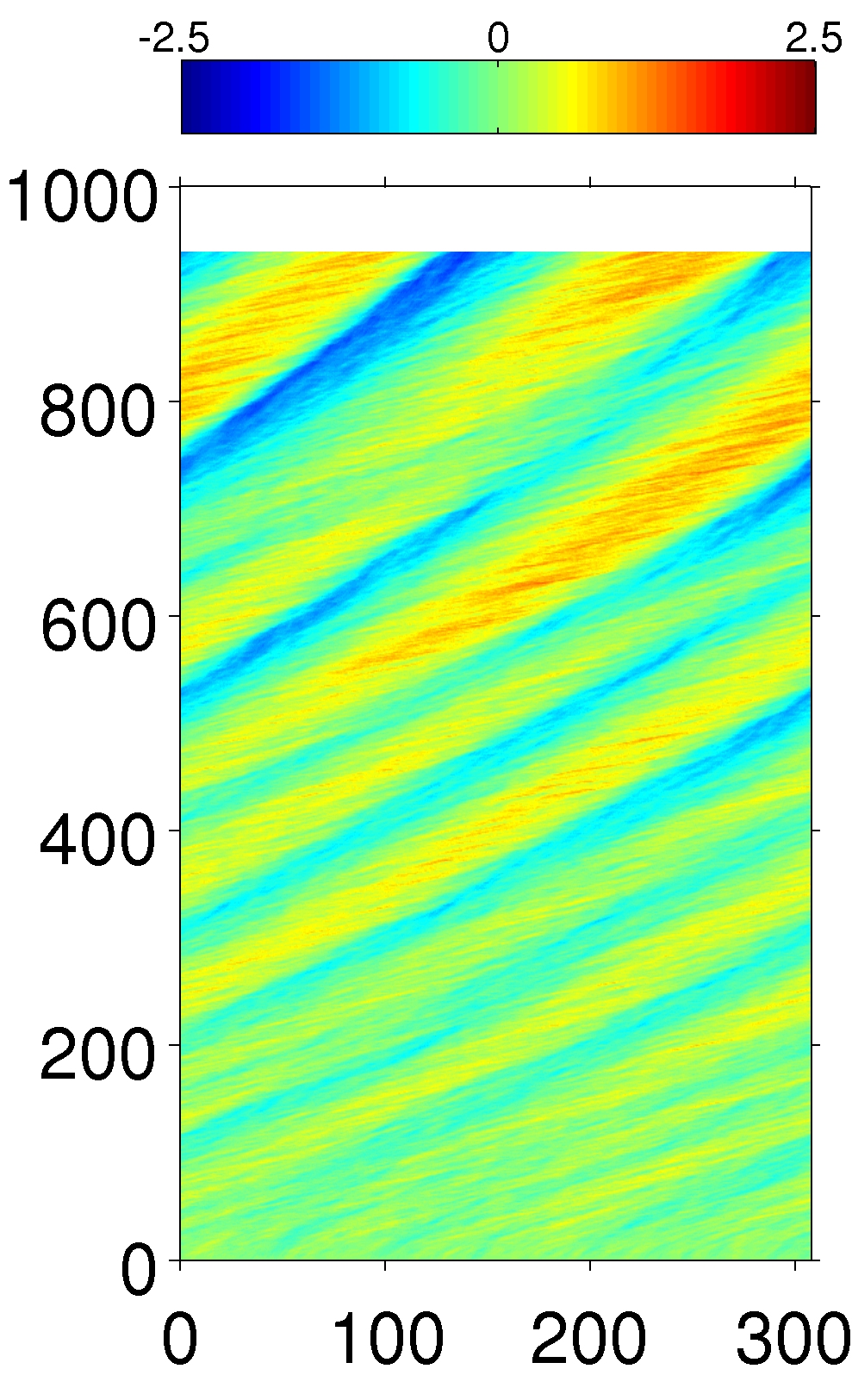}
          \centerline{\small $x/\Dia$ }
        \end{minipage}
        \hspace{1ex}
        \raisebox{20ex}{$(c)$}
        \begin{minipage}{2ex}
          \rotatebox{90}
          {\small $t \, \ubulk/\hmean$}
        \end{minipage}
        \begin{minipage}{0.25\linewidth}
          \includegraphics[width=\linewidth]
          {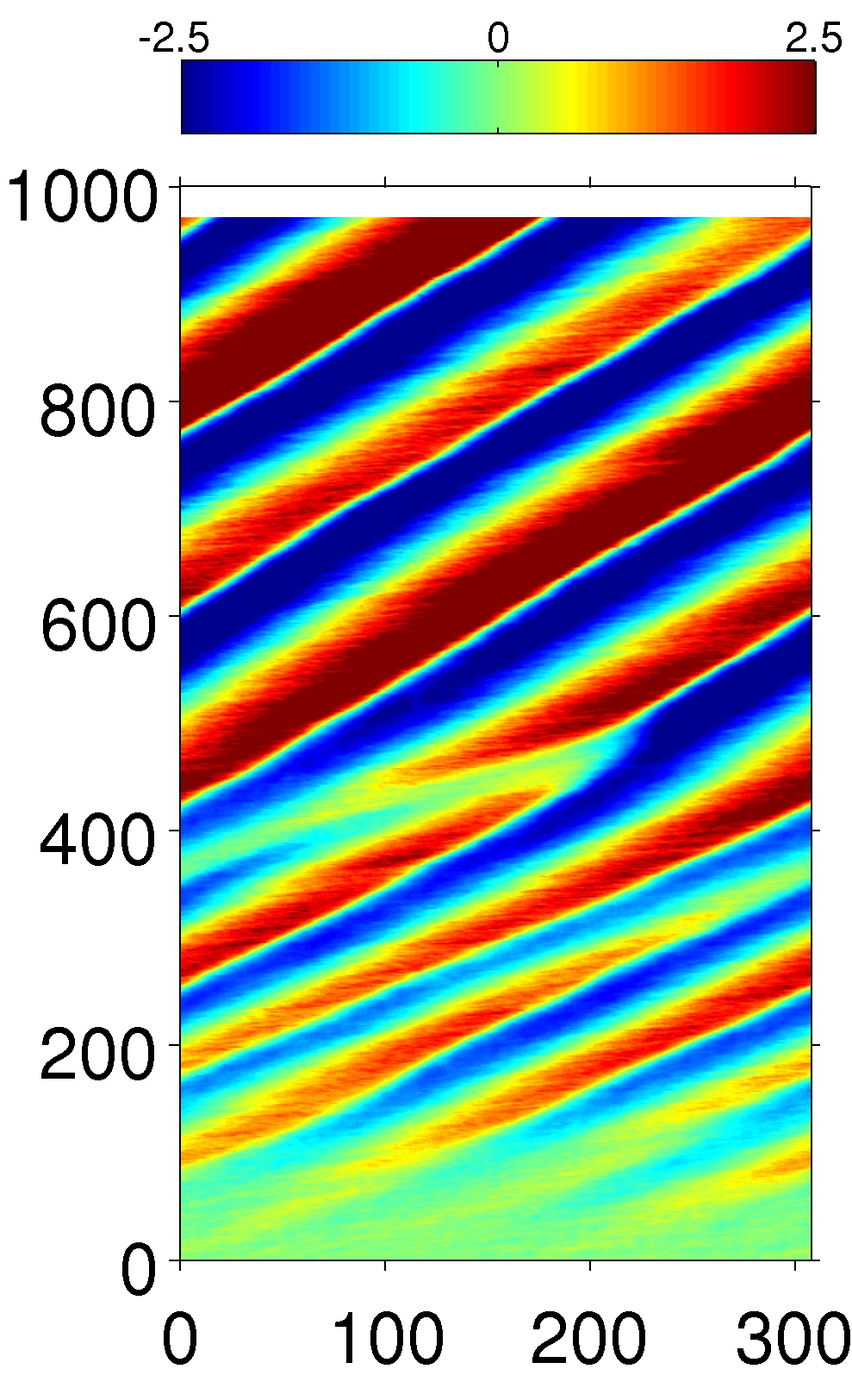}
          \centerline{\small $x/\Dia$ }
        \end{minipage}
        \caption{Space-time evolution of the fluctuation of the
          fluid-bed interface location,
          $\hbedzmean^\prime(x,t)=\hbedzmean(x,t)-\hbedmean(t)$, 
          normalized with the particle diameter $D$: 
          (\textit{a})~case \caseLa;   
          (\textit{b})~case \caseLb; 
          (\textit{c})~case \caseTa. 
          %
        }
        \label{fig:space-time-plot-of-bed-interface}
\end{figure}
\subsection{Definition of the fluid-bed interface}
\label{determination-of-the-fluid-bed-interface}
The location of the interface between the fluid and the sediment bed
has been determined in the following way. 
First, a solid phase indicator function $\phi_p(\mathbf{x},t)$ is
defined which measures unity if $\mathbf{x}$ is located inside any
particle and zero elsewhere. 
Spanwise averaging then yields 
$\phimeansmoothspanwise(x,y,t)$ which is a direct measure of the 
instantaneous, two-dimensional solid volume fraction. 
The spanwise-averaged fluid-bed interface location $\hbedzmean(x,t)$ is
finally extracted by means of a threshold value, chosen as
$\phimeansmoothspanwise^{thresh} = 0.1$ 
\citep{kidanemariam:14a}, 
viz.
%
%
\begin{eqnarray}\label{eq:fluid-height-and-bed-height} 
   \hbedzmean(x,t)   &=& y \;\; \vert \; \phimeansmoothspanwise(x,y,t)
                                =\phimeansmoothspanwise^{thresh}\;,
\end{eqnarray}
%
as illustrated in figure~\ref{fig:bed_thickness_evolution}\textit{(b)}. 
The corresponding spanwise-averaged fluid height is then simply given
by $\hfluidzmean(x,t)=\Ly - \hbedzmean(x,t)$.  
Figure~\ref{fig:bed_thickness_evolution}\textit{(c,d)} shows the time
evolution of 
the streamwise average of the bed height defined in
(\ref{eq:fluid-height-and-bed-height}), $\hbedmean(t)$, 
as well as its minimum and maximum values. 
%
%
It can be observed that after a small initial dilation \hbedmean\
quickly reaches an approximately constant value in all three cases. 
Contrarily, the maximum and minimum values of the bed elevation 
continue to diverge over the simulated interval of approximately
$1000$ bulk time units, not reaching an equilibrium state.  
%

%
%
\section{Results}
\label{sec:results-and-discussion}
Space-time plots of the sediment bed height fluctuation with respect
to the instantaneous streamwise average, defined as 
$\hbedzmean^\prime(x,t)=\hbedzmean(x,t)-\hbedmean(t)$, 
are shown in figure~\ref{fig:space-time-plot-of-bed-interface}. 
The emergence of dune-like patterns can be clearly observed, with a
streamwise succession of alternating ridges and troughs forming
shortly after start-up in all three cases. 
%
%
In the two simulations in the laminar regime we obtain similar
fluctuation amplitudes.  
However, the propagation velocity is significantly larger in case~\caseLb\
(i.e.\ at larger Shields number) than in case~\caseLa. 
The turbulent case \caseTa, on the other hand, is found to exhibit a
higher growth rate, rapidly leading to enhanced fluctuation
amplitudes as compared to both laminar cases. 
\begin{figure}
  \begin{minipage}{2.5ex}
    \rotatebox{90}{\small $y/D$}
  \end{minipage}
  \begin{minipage}{0.48\linewidth}
    \centerline{ $(a)$}
    \includegraphics[width=\linewidth]
    {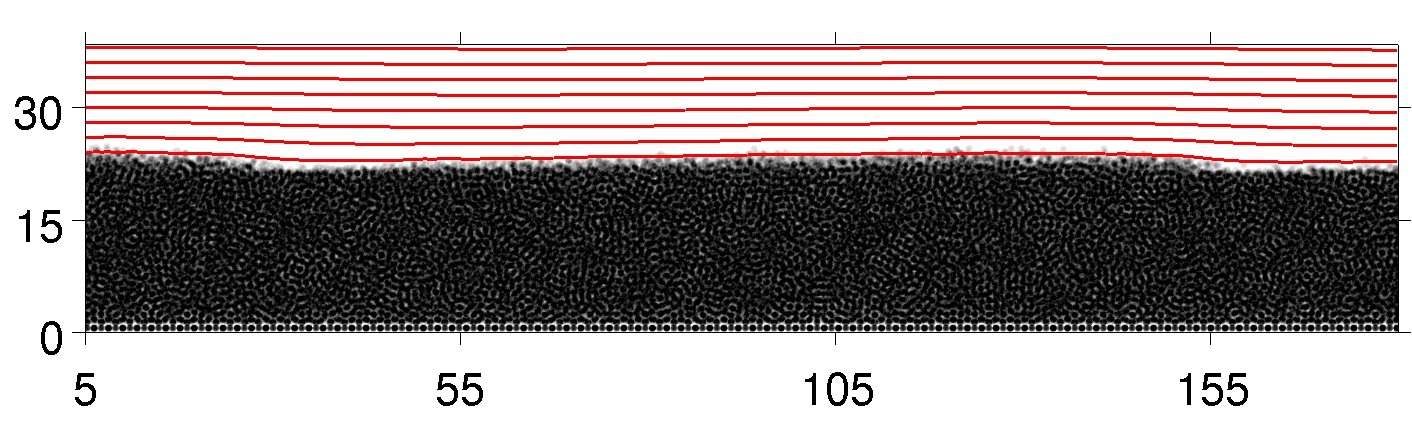}
    \centerline{ $x/D$}
  \end{minipage}
  \begin{minipage}{0.48\linewidth}
    \centerline{ $(b)$}
    \includegraphics[width=\linewidth]
    {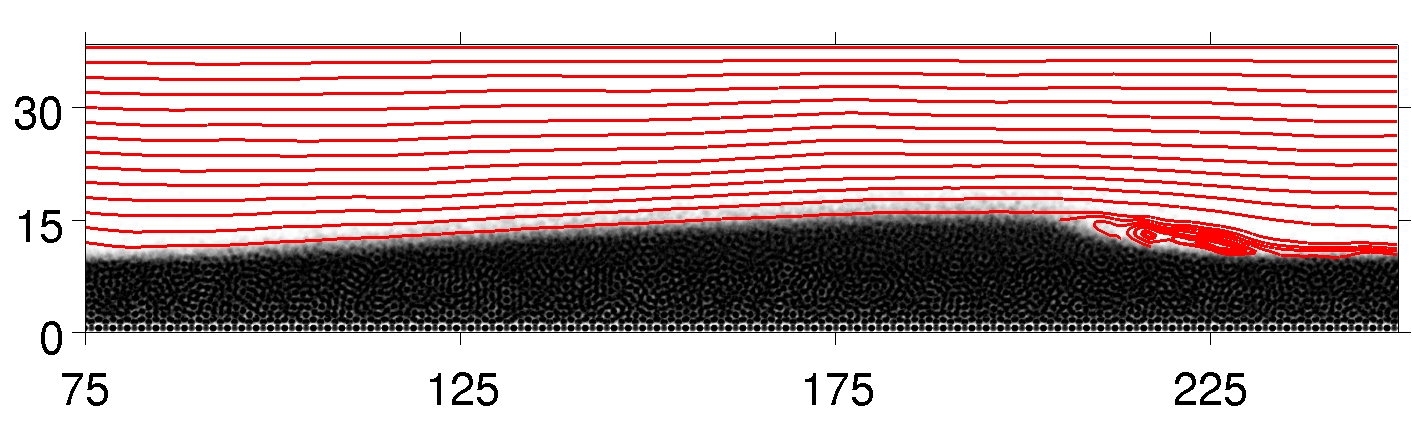}
    \centerline{ $x/D$}
  \end{minipage}
  \caption{
    Close-up of the instantaneous,
    spanwise-averaged  
    solid-volume fraction \phimeansmoothspanwise\ (plotted in
    greyscale)  
    as well as the 
    streamlines computed from a spanwise-averaged
    instantaneous flow field in the final phase of the simulated
    interval:   
    $(a)$ case~\caseLa; 
    $(b)$ case~\caseTa. 
  }
  \label{fig:instantaneous-spanwise-average-streamlines}
\end{figure}
Furthermore, these space-time plots 
show the
occasional occurrence of dune mergers with a subsequent increase of
wavelength and an apparent decrease of the propagation speed. 
For times $t\gtrsim550\hmean/\ubulk$ the sediment bed patterns in the turbulent case \caseTa\ 
(cf.\ figure~\ref{fig:space-time-plot-of-bed-interface}$c$) 
remain roughly invariant with two distinct dunes featuring somewhat
different elevation amplitudes.    

A visualization of the fluid-bed interface and the streamlines of the
spanwise averaged flow field towards the end of the
simulated intervals is shown in
figure~\ref{fig:instantaneous-spanwise-average-streamlines}. 
It is found that the patterns in the laminar cases indeed
correspond to `small dunes' in the terminology of
\citet{Ouriemi2009b}, and to `vortex dunes' with significant
separation on the lee-side in the turbulent case (the
graph for case~\caseLb\ is similar to case~\caseLa\ and has been
omitted). These results are, therefore, consistent with the regime
classification based upon the channel (or pipe) Reynolds number
proposed by these authors (cf.\ figure~\ref{fig:parameter-space}). 

\begin{figure}
  \centering
  \raisebox{12ex}{$(a)$} 
  \begin{minipage}{2ex}
    \rotatebox{90}
    {\small $\lambda_{av}/\Dia$}
  \end{minipage}
  \begin{minipage}{.43\linewidth}
    \includegraphics[width=\linewidth]
    {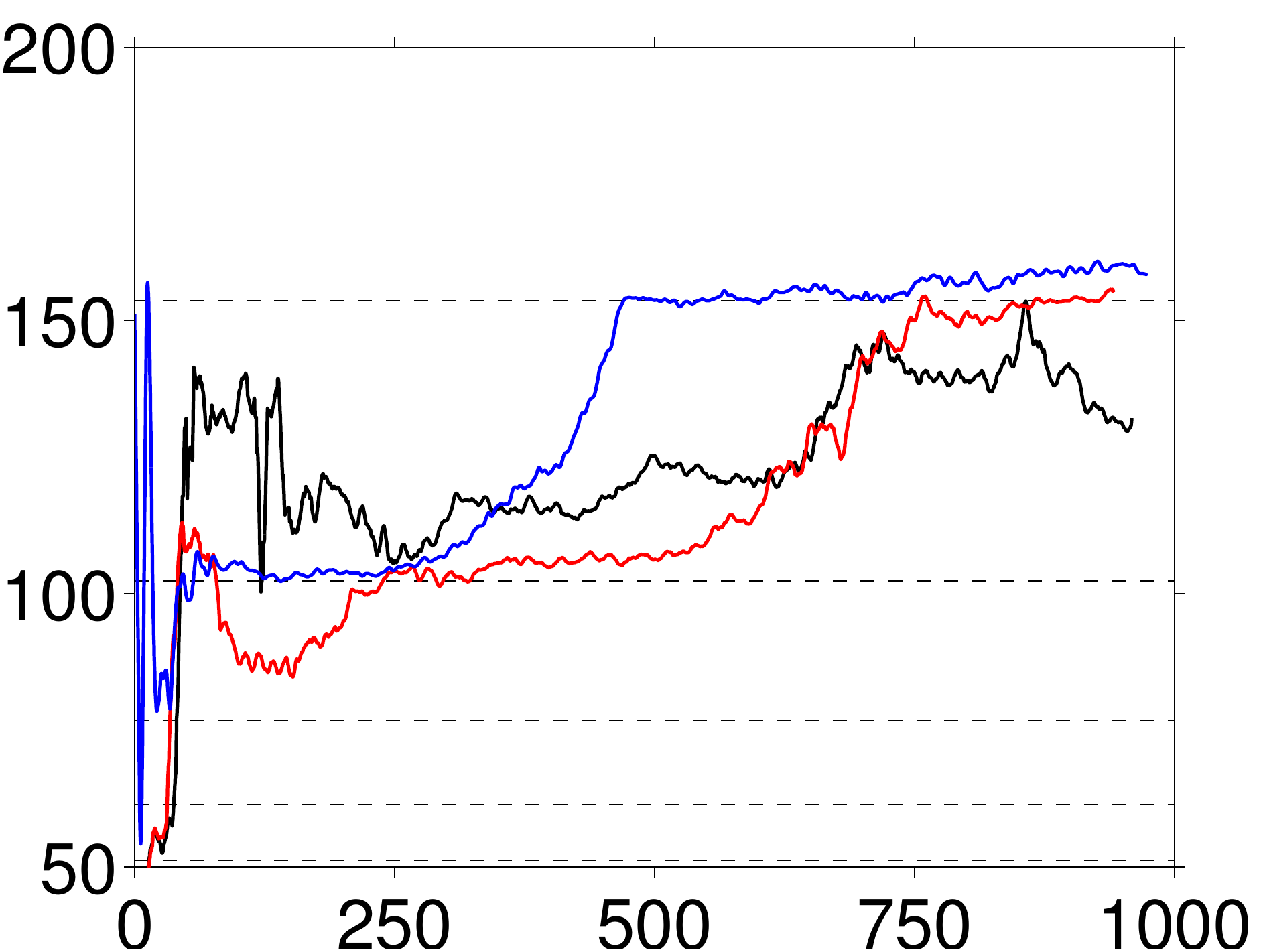}
    \centerline{\small
      $t\, \ubulk/\hmean$}
  \end{minipage}
  \raisebox{12ex}{$(b)$} 
  \begin{minipage}{3ex}
    \rotatebox{90}
    {\small $\rmsamplitude/\Dia$}
  \end{minipage}
  \begin{minipage}{.4\linewidth}
    \includegraphics[width=\linewidth]
    {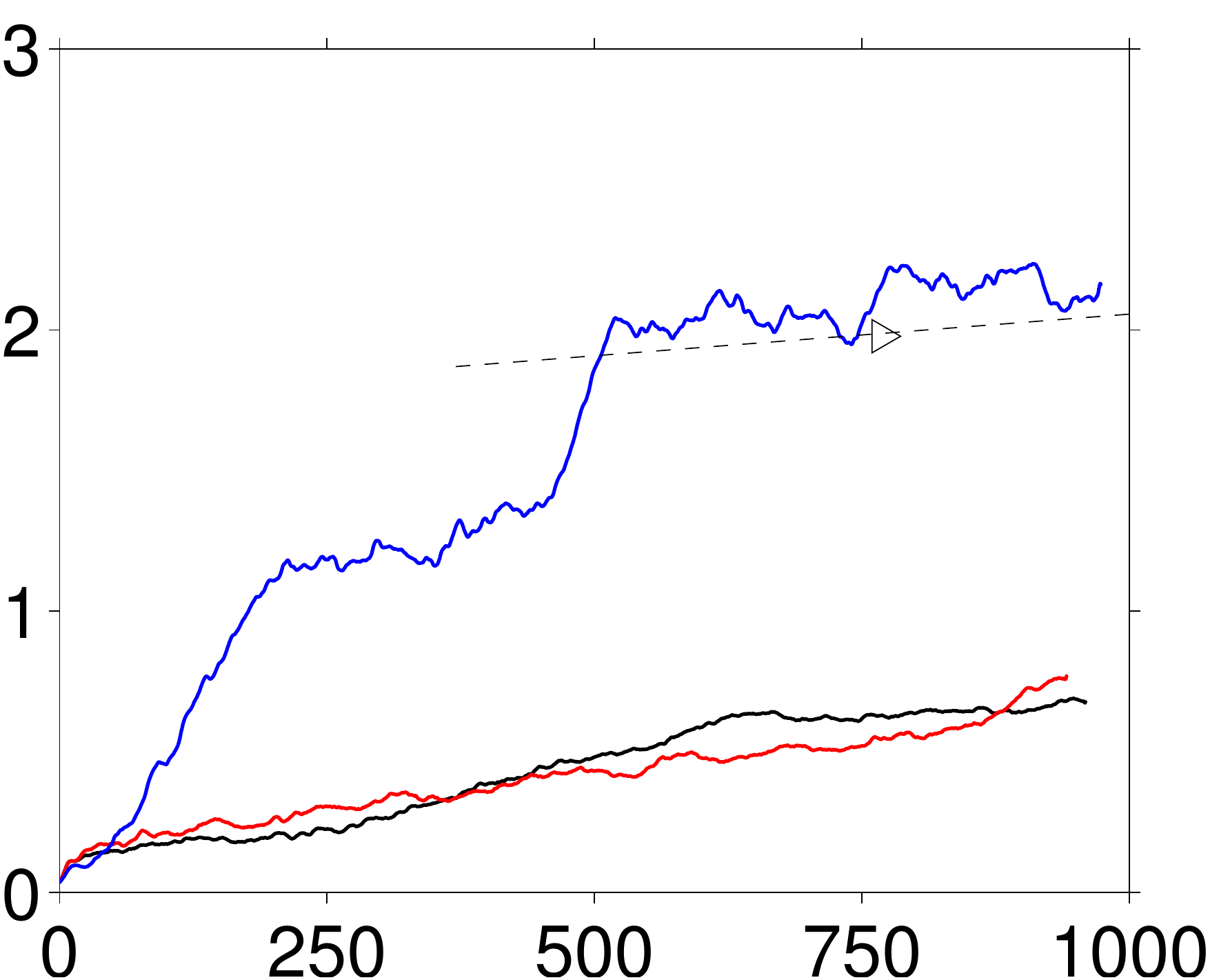}
    \centerline{\small
      $t\, \ubulk/\hmean$}
  \end{minipage}
  \\
  \caption{
    (\textit{a}) 
    Time evolution of the mean wavelength
    of the sediment bed height normalized with the particle
    diameter. The dashed lines indicate the wavelengths of the second
    to sixth streamwise harmonics in the current domain.
    (\textit{b}) 
    Time evolution of the r.m.s.\ sediment bed
    height. The dashed line shows the fit obtained by
    \citet[figure~6$a$]{Langlois2007a} for their case with
    $Re_{2\hmean}=15130$, 
    $\shields=0.099$, 
    $\hmean/\Dia=35$. 
    Note that this fit was obtained for $t\geq740\ubulk/\hmean$ 
    (their first data point is indicated by the symbol
    `$\vartriangleright$').  
    In both graphs solid lines with the following colors correspond to
    the present cases:   
    {\color{black}\solidthick}~\caseLa, 
    {\color{red}\solidthick}~\caseLb, 
    {\color{blue}\solidthick}~\caseTa.
  }
  \label{fig:time-evolution-wavelength-amplitude}
\end{figure}

%
The instantaneous two-point correlation of the bed
height fluctuation as a function of streamwise separations $r_x$,
defined as
$R_{h}(r_x,t)=\langle\hbedzmean^\prime(x,t)\,\hbedzmean^\prime(x+r_x,t)\rangle_x$,  
exhibits a clear negative minimum in all of the present cases
(figure omitted). 
Therefore, we can define an average pattern wavelength $\meanlambda$ as twice 
the streamwise separation at which the global minimum of $R_{h}$
occurs. 
%
The time evolution of the mean wavelength $\meanlambda$,
normalized by the particle diameter, is shown in
figure~\ref{fig:time-evolution-wavelength-amplitude}$(a)$. 
Also indicated with horizontal dashed lines in the graph are the
wavelengths of the second to sixth harmonics 
in the present computational domain (recall that $L_x/D=307.2$
throughout the present work). 
It can be seen that for short times the wavelength $\meanlambda$ in
the turbulent case~\caseTa\ exhibits several fast oscillations between $L_x/5$
and $L_x/2$ before approximately settling at a value near $L_x/3$ (i.e.\
$\meanlambda/\Dia\approx102$) for some 200 bulk time units. Starting
with time $t\ubulk/\hmean\approx250$ the average wavelength then grows
at an increasing rate, settling again at
$\meanlambda\approx L_x/2=153.6\Dia$ until the end of the simulated
interval. 
%
%
%
Contrarily, the two laminar cases have a less oscillatory initial
evolution. Case~\caseLb\ first settles into a plateau-like state (with
$\meanlambda\approx L_x/3$) after approximately 250 elapsed bulk time
units. 
Subsequently the wavelength corresponding to
the second harmonic ($\meanlambda\approx L_x/2$) 
grows in amplitude and becomes dominant 
after approximately 750 bulk units. 
Case~\caseLa\ does not appear to settle into any of the harmonic
wavelengths of the current domain, exhibiting an average wavelength in
the range of $100-150$ particle diameters for the most part of the
simulated interval.  
In turbulent channel flow experiments, \citet{Langlois2007a} have
determined values of the initial pattern wavelength of
$\lambda/D\approx100-150$, roughly independent of the grain size. 
%
Upon scaling with the equivalent boundary layer thickness, the average
pattern wavelengths in the three cases of the present work measure
$3-6.5\hmean$, except for a few initial bulk time units. This range is
comparable to the range found for the initial wavelength of the `small
dunes' in pipe flow reported as $2.5-12.6\hmean$ by
\citet[figure~7$b$]{Ouriemi2009b} and for the `vortex dune' data shown
in their figure~3$(c)$, where an initial wavelength of $4\hmean$ is
observed. 
%

%
Let us turn to the amplitude of the sediment patterns. 
The evolution of the r.m.s.\ value of the fluid-bed interface location  
is plotted in
figure~\ref{fig:time-evolution-wavelength-amplitude}(\textit{b}). 
The fact that no saturation is observed by the end of the simulated
intervals is consistent with experimental observations, where it was
found that even after an order of magnitude longer times the amplitude
of the patterns continues to grow \citep[the intervals simulated in the
current work correspond to roughly one minute in the experiments of]
[conducted over more than one hour]{Langlois2007a}. 
Both of the present laminar cases exhibit growth at an approximately
constant rate (with a slope of $6\cdot10^{-4}$ in the units of
figure~\ref{fig:time-evolution-wavelength-amplitude}\textit{b}). 
Contrarily, the time evolution in the turbulent case~\caseTa\ is
quasi-linear with different slopes in different time intervals. 
The initial growth of the turbulent case (for times up to
$t\ubulk/\hmean\approx200$) and the growth in the interval
$450\lesssim t\ubulk/\hmean\lesssim500$ 
are approximately ten times higher than the growth in the
laminar cases, while in the remaining two intervals the growth rate is
comparable to the laminar value. 
As can be seen in
figure~\ref{fig:time-evolution-wavelength-amplitude}(\textit{b}), the
time evolution in the final period 
of case~\caseTa\ is roughly equivalent to the one determined by
\citet{Langlois2007a} in turbulent flow at comparable parameter values 
($Re_{2\hmean}=15130$, $\shields=0.099$, $\hmean/\Dia=35$). 

The propagation speed of the
patterns can be determined from the shift of the maximum of the
two-point/two-time correlation of the fluid-bed interface
fluctuation $\hbedzmean^\prime(x,t)$.
%
It turns out that the patterns in case~\caseLa\ propagate at a
relatively constant speed of approximately $0.011u_b$, while the
propagation velocity decreases with time during the coarsening process
in cases~\caseLb\ and \caseTa, reaching values of $0.026u_b$ and
$0.035u_b$, respectively, in the final period of the current
simulations. 
The latter number is consistent with the range of values reported 
for `vortex dunes' by \citet[figure~3$b$]{Ouriemi2009b}. 

  \begin{figure}
    \raisebox{3.5ex}{
      \begin{minipage}{2ex}
        $(a)$\\[2ex]
        \rotatebox{90}{$\langle\overline{q}_p\rangle_x/q_v$}
        \\[4ex]
        $(b)$\\[2ex]
        \rotatebox{90}{$\langle\overline{q}_p\rangle_x/q_i$}
      \end{minipage}
    }
    \begin{minipage}{.45\linewidth}
      \includegraphics[width=\linewidth]
      {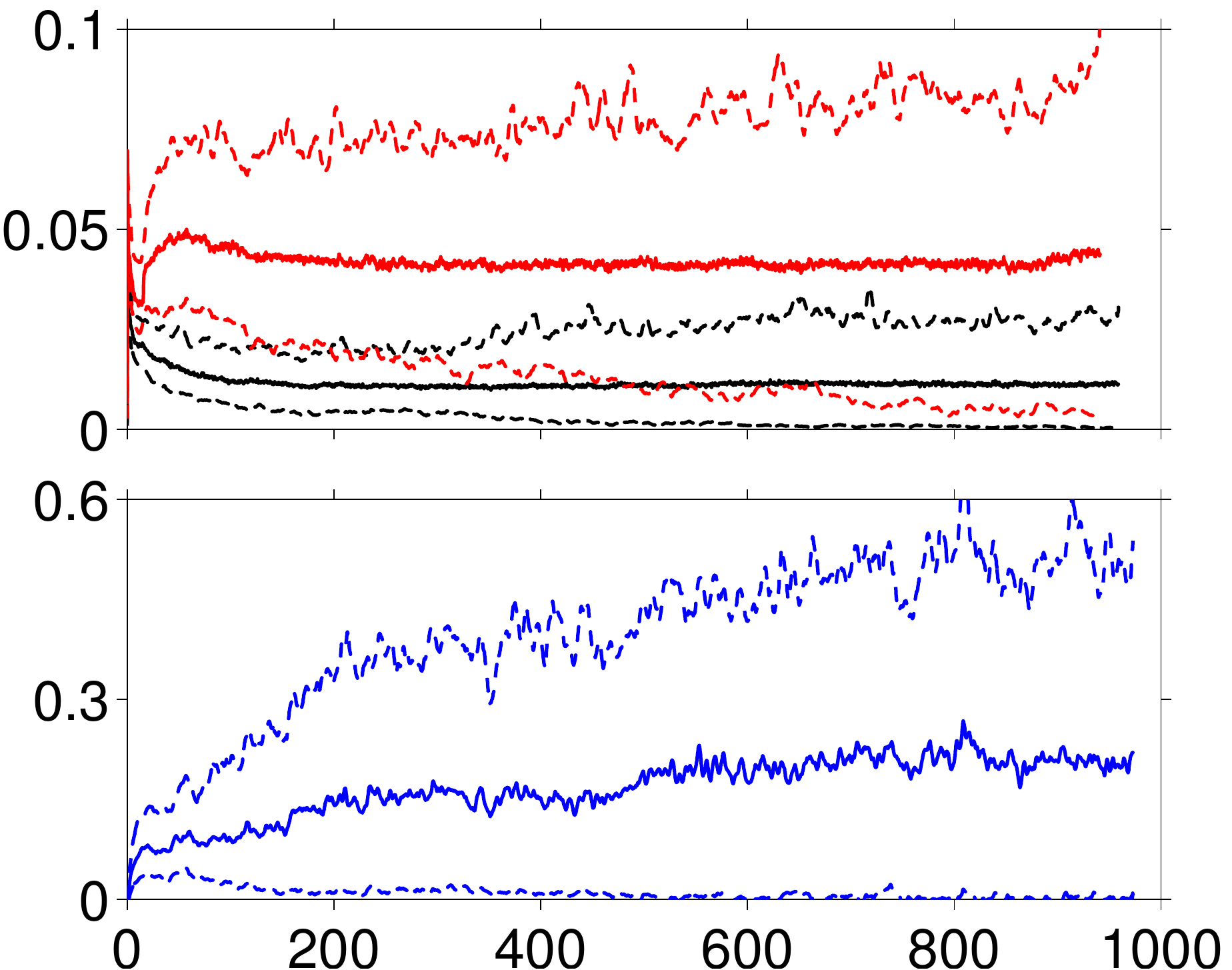}
      \centerline{\small         $t\,\ubulk/\hmean$}
    \end{minipage}
    \hfill
    \raisebox{5ex}{
      \begin{minipage}{2ex}
        $(c)$\\[3ex]
        \rotatebox{90}{$\langle\overline{q}_p\rangle_{xt}/q_i$}
        \\[2ex]
        \rotatebox{90}{$\langle\overline{q}_p\rangle_{xt}/q_v$,}
      \end{minipage}
    }
    \begin{minipage}{.45\linewidth}
      \includegraphics[width=\linewidth]
      {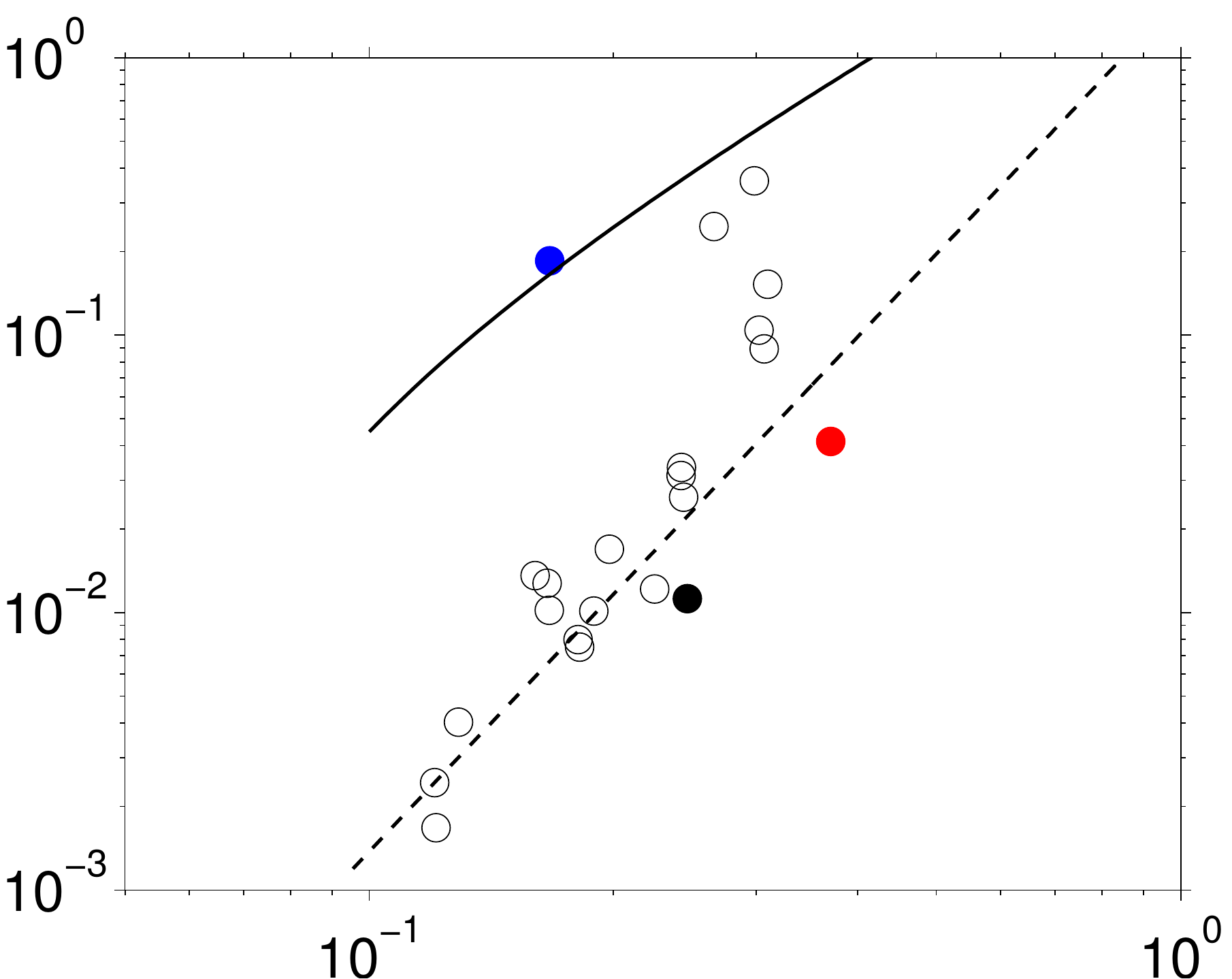}
      \centerline{\small
        $\shields$}
    \end{minipage}
    \caption{
        Analysis of the volumetric particle flow rate 
        (per unit spanwise length), $\overline{q}_p(x,t)$.
        (\textit{a})~The time evolution of the streamwise average value, 
        $\langle\overline{q}_p\rangle_x(t)$ 
        in cases \caseLa\ (black color) and
        \caseLb\ (red color) is shown with solid lines. 
        The viscous scale $q_v=Ga^2\,\nu$ is used for the purpose of
        normalization. 
        Additionally, the two dashed lines in each case indicate the respective
        minimum and maximum values ($\min_x\overline{q}_p$,
        $\max_x\overline{q}_p$). 
        (\textit{b})~The same quantity in case \caseTa, using the
        inertial scale $q_i=u_gD$ for normalization.
        (\textit{c})~The value of the time average
        $\langle\overline{q}_p\rangle_{xt}$ 
        over the final part of the simulations plotted versus the
        Shields number $\shields$: 
        {\color{black}\solidcircle},~\caseLa;
        {\color{red}\solidcircle},~\caseLb;
        {\color{blue}\solidcircle},~\caseTa. 
        Note the different scalings ($q_v$ in the laminar cases, $q_i$ in
        the turbulent case). 
        The open circles are for featureless bedload transport in
        laminar flow 
        \citep{kidanemariam:14a}; 
        the dashed line is
        the fit
        $\langle\overline{q}_p\rangle_{xt}/q_v=1.66\,\shields^{3.08}$ from
        that reference.   
        The solid line is the \cite{wong:06} version 
        $\langle\overline{q}_p\rangle_{xt}/q_i=4.93\,(\shields-0.047)^{1.6}$ 
        of the \cite{meyer-peter:48} formula for turbulent flow.
    }
    \label{fig-total-particle-flow-rate}
  \end{figure}
  %
%
  %
  The volumetric particle flow rate (per unit spanwise length),
  $\overline{q}_p(x,t)$, is analyzed in
  figure~\ref{fig-total-particle-flow-rate}. 
  The solid lines in 
  figure~\ref{fig-total-particle-flow-rate}$(a,b)$ show the temporal
  evolution of the streamwise average 
  $\langle\overline{q}_p\rangle_x(t)$, which is observed to reach 
  approximately constant values after a few hundred bulk time units in
  all cases. The continuous growth of the pattern amplitudes (cf.\
  figure~\ref{fig:time-evolution-wavelength-amplitude}$b$) seems to
  have only a mild influence upon the total particle flow rate,
  irrespective of the flow regime. 
  These graphs also show for each instant the maximum and minimum
  values (in space) of the particle flow rate, drawn as dashed
  lines. 
  Although these extrema curves are noisier, it can be observed
  that the maxima continue to grow until the end of the simulations,
  consistent with the increase in the amplitude of the propagating
  patterns. 
  Of particular interest in view of applications is the scaling of the
  particle transport rates, typically expressed as a function of the
  Shields number $\shields$. 
  Figure~\ref{fig-total-particle-flow-rate}$(c)$ shows the
  space-averaged values, additionally averaged in time over the final 
  part of the simulations, 
  denoted as 
  $\langle\overline{q}_p\rangle_{xt}$.  
  It is found that the present values of
  $\langle\overline{q}_p\rangle_{xt}/q_v$ (where $q_v=Ga^2\,\nu$) in
  the two laminar cases are only slightly below the (approximately)
  cubic power law fitted by 
  \cite{kidanemariam:14a} 
  to their
  simulation data for featureless bedload transport.\footnote{%
      Note that in 
      \citep{kidanemariam:14a} 
      the Shields number
      (termed $\shields_{Pois}$ therein) was defined based upon the
      assumption of a parabolic fluid velocity profile for consistency
      with the reference experiment.  
      As a result, the fit represents even the data points at larger
      values of the Shields number $\shields_{Pois}$ with good accuracy.
  }
  It is obviously not possible to infer scaling from two data points. 
  However, if a power law of the particle flow rate as a function of
  the Shields number is assumed, the present laminar data suggest a
  cubic variation.  
  Turning to the turbulent case \caseTa,
  figure~\ref{fig-total-particle-flow-rate}$(c)$ shows that the value for
  $\langle\overline{q}_p\rangle_{xt}/q_i$ (with the inertial scaling
  $q_i=u_g\,D$) is very close to the value given by the empirical law
  of \cite{wong:06}, which in turn is a modified version of the
  \cite{meyer-peter:48} formula for turbulent flow. 
  Wong \& Parker's formula is valid for plane sediment beds. 
  The fact that the present data agrees well with that prediction
  together with the observed mild variation in time 
  (cf.\ figure~\ref{fig-total-particle-flow-rate}$b$) 
  shows that the presence of `vortex dunes' does not strongly affect
  the net particle transport rate. 
  %
%

%
%
\section{Summary and conclusion}
\label{sec:conclusion}
We have performed direct numerical simulation of the flow over an
erodible bed of spherical sediment particles above both the threshold
for particle mobility and for pattern formation. 
Two cases in laminar flow (with different Galileo and Shields numbers)
lead to the formation of `small dunes', 
while one case under turbulent flow conditions exhibits `vortex dunes', 
consistently with the regime classification of \citet{Ouriemi2009b}. 
The reconstruction of the fluid-bed interface from a spanwise-averaged
solid volume fraction (involving a threshold value) has allowed us to
analyze the length scales, amplitude and propagation velocity of the
sediment patterns in detail. 
In all three respects, the results of the present simulations are found
to be consistent with available experimental data. 
  We have observed that the continuing growth of the dune patterns,
  which have not reached a statistically stationary state after
  approximately 1000 bulk time units, does not strongly affect the net
  volumetric particle transport rates. 
  In the two laminar cases the particle flow rate (per unit span) is
  consistent with a cubic power law as a function of the Shields
  number; these values are found to be not far from those obtained in
  featureless bedload transport. 
  The value pertaining to the turbulent case is very well predicted by
  the transport law of \cite{wong:06} which is derived for turbulent
  flow in the presence of a plane mobile bed. 
  The present results therefore seem to suggest that the presence of
  `small dunes' as well as that of `vortex dunes' up to the amplitudes
  encountered in the present simulations does not lead to a
  modification of the net particle transport rate which would require
  a correction of the respective transport laws. 
  This conclusion should be reassessed in the future when much longer
  temporal intervals can be covered. 

  The present work demonstrates that the DNS-DEM approach to sediment
  pattern formation is feasible today. Although still costly in terms of
  computational resources, it is already possible to address some of
  the outstanding questions with this method. 
  Some aspects which are of importance in geophysical
  applications (such as reaching the fully rough turbulent regime, 
  guaranteeing an asymptotically large computational domain and
  integrating over asymptotically long temporal intervals) 
  still present 
  a considerable computational challenge. 
  %
  
  As a next step, the streamwise length of the computational domain
  should be extended in order to reduce the influence of the
  discreteness of the numerical harmonics upon the pattern 
  wavelength. 
  Conversely, shrinking the box length will allow to reveal the
  smallest amplified wavelength of the sedimentary patterns. 
  Finally, an in-depth investigation of the flow field which develops
  over the time-dependent sediment bed can be carried out based upon
  the simulation data. 
  Preliminary visualization suggests that in the
  turbulent case the coherent structures leave their footprint in the
  bed shape, visible as longitudinal ridges and troughs
  superposed on the roughly two-dimensional dune patterns. 
  Such an analysis is left for future work.
%
%

%
%

\vspace*{1ex}
%
%
%
This work was supported by the German Research Foundation (DFG)
through grant {UH~242/2-1}. 
%
The computer resources, technical expertise and assistance provided by
the staff at LRZ M\"unchen (grant pr58do) are thankfully
acknowledged. 
%

%

\vspace*{1ex}
%
  Movies of the particle motion are available as 
  ancillary files along with the arXiv submission         
  and from the following URL:\\
\centerline{\href{http://www.ifh.kit.edu/dns_data}
{\tt http://www.ifh.kit.edu/dns\_data}}

%
\bibliographystyle{model2-names}
\addcontentsline{toc}{section}{References}

\begin{thebibliography}{25}
\expandafter\ifx\csname natexlab\endcsname\relax\def\natexlab#1{#1}\fi

\bibitem[Betat {\em et~al.\/}(2002)Betat, Kruelle, Frette \&
  Rehberg]{Betat2002a}
{\sc Betat, A., Kruelle, C.~A., Frette, V. \& Rehberg, I.} 2002 {Long-time
  behavior of sand ripples induced by water shear flow.} {\em Eur. Phys. J. E.
  Soft Matter\/} {\bf 8}~(5), 465--76.

\bibitem[Chan-Braun {\em et~al.\/}(2011)Chan-Braun, Garc\'{\i}a-Villalba \&
  Uhlmann]{Chan-braun2011}
{\sc Chan-Braun, C., Garc\'{\i}a-Villalba, M. \& Uhlmann, M.} 2011 {Force and
  torque acting on particles in a transitionally rough open-channel flow}. {\em
  J. Fluid Mech.\/} {\bf 684}, 441--474.

\bibitem[Charru(2006)]{Charru2006c}
{\sc Charru, F.} 2006 {Selection of the ripple length on a granular bed sheared
  by a liquid flow}. {\em Phys. Fluids\/} {\bf 18}~(12), 121508.

\bibitem[Charru \& Hinch(2006)]{Charru2006d}
{\sc Charru, F. \& Hinch, E.~J.} 2006 {Ripple formation on a particle bed
  sheared by a viscous liquid. Part 1. Steady flow}. {\em J. Fluid Mech.\/}
  {\bf 550}, 111--121.

\bibitem[Charru \& Mouilleron-Arnould(2002)]{Charru2002}
{\sc Charru, F. \& Mouilleron-Arnould, H.} 2002 {Instability of a bed of
  particles sheared by a viscous flow}. {\em J. Fluid Mech.\/} {\bf 452},
  303--323.

\bibitem[Coleman {\em et~al.\/}(2003)Coleman, Fedele \& Garcia]{Coleman2003}
{\sc Coleman, S.~E., Fedele, J.~J. \& Garcia, M.~H.} 2003 {Closed-Conduit
  Bed-Form Initiation and Development}. {\em J. Hydraul. Eng.\/} {\bf
  129}~(December), 956--965.

\bibitem[Coleman \& Melville(1994)]{Coleman1994}
{\sc Coleman, S.~E. \& Melville, B.~W.} 1994 {Bed-form development}. {\em J.
  Hydraul. Eng.\/} {\bf 120}~(4), 544--560.

\bibitem[Coleman \& Nikora(2009)]{Coleman2009}
{\sc Coleman, S.~E. \& Nikora, V.~I.} 2009 {Bed and flow dynamics leading to
  sediment-wave initiation}. {\em Water Resour. Res.\/} {\bf 45}~(4), n/a--n/a.

\bibitem[Colombini(2004)]{Colombini2004}
{\sc Colombini, M.} 2004 {Revisiting the linear theory of sand dune formation}.
  {\em J. Fluid Mech.\/} {\bf 502}, 1--16.

\bibitem[Colombini \& Stocchino(2011)]{Colombini2011}
{\sc Colombini, M. \& Stocchino, a.} 2011 {Ripple and dune formation in
  rivers}. {\em J. Fluid Mech.\/} {\bf 673}, 121--131.

\bibitem[Engelund \& Fredsoe(1982)]{Engelund1982}
{\sc Engelund, F. \& Fredsoe, J.} 1982 {Sediment Ripples and Dunes}. {\em Annu.
  Rev. Fluid Mech.\/} {\bf 14}~(1), 13--37.

\bibitem[Garc\'{\i}a-Villalba {\em et~al.\/}(2012)Garc\'{\i}a-Villalba,
  Kidanemariam \& Uhlmann]{Garcia-villalba2012}
{\sc Garc\'{\i}a-Villalba, M., Kidanemariam, A.~G \& Uhlmann, M.} 2012 {DNS of
  vertical plane channel flow with finite-size particles: Voronoi analysis,
  acceleration statistics and particle-conditioned averaging}. {\em Int. J.
  Multiph. Flow\/} {\bf 46}, 54--74.

\bibitem[Kennedy(1963)]{Kennedy1963}
{\sc Kennedy, J.~F.} 1963 {The mechanics of dunes and antidunes in erodible-bed
  channels}. {\em J. Fluid Mech.\/} {\bf 16}~(4), 521--544.

\bibitem[Kidanemariam \& Uhlmann(2014)]{kidanemariam:14a}
{\sc Kidanemariam, A.G. \& Uhlmann, M.} 2014 Interface-resolved direct
  numerical simulation of the erosion of a sediment bed sheared by laminar
  flow. {\em Int.\ J.\ Multiphase Flow\/} {\it (submitted)}.

\bibitem[Kidanemariam {\em et~al.\/}(2013)Kidanemariam, Chan-Braun, Doychev \&
  Uhlmann]{Kidanemariam2013}
{\sc Kidanemariam, A.~G., Chan-Braun, C., Doychev, T. \& Uhlmann, M.} 2013
  {Direct numerical simulation of horizontal open channel flow with
  finite-size, heavy particles at low solid volume fraction}. {\em New J.
  Phys.\/} {\bf 15}~(2), 025031.

\bibitem[Langlois \& Valance(2007)]{Langlois2007a}
{\sc Langlois, V. \& Valance, A.} 2007 {Initiation and evolution of current
  ripples on a flat sand bed under turbulent water flow.} {\em Eur. Phys. J. E.
  Soft Matter\/} {\bf 22}~(3), 201--8.

\bibitem[Meyer-Peter \& M\"uller(1948)]{meyer-peter:48}
{\sc Meyer-Peter, E. \& M\"uller, R.} 1948 Formulas for bed-load transport. In
  {\em Proc. 2nd Meeting\/}, pp. 39--64. IAHR, Stockholm, Sweden.

\bibitem[Ouriemi {\em et~al.\/}(2009)Ouriemi, Aussillous \&
  Guazzelli]{Ouriemi2009b}
{\sc Ouriemi, M., Aussillous, P. \& Guazzelli, \'{E}.} 2009 {Sediment dynamics.
  Part 2. Dune formation in pipe flow}. {\em J. Fluid Mech.\/} {\bf 636},
  295--319.

\bibitem[Raudkivi(1997)]{Raudkivi1997}
{\sc Raudkivi, A.~J.} 1997 {Ripples on Stream Bed}. {\em J. Hydraul. Eng.\/}
  {\bf 123}~(1), 58--64.

\bibitem[Richards(1980)]{Richards1980}
{\sc Richards, K.~J.} 1980 {The formation of ripples and dunes on an erodible
  bed}. {\em J. Fluid Mech.\/} {\bf 99}~(3), 597--618.

\bibitem[Sumer \& Bakioglu(1984)]{Sumer1984}
{\sc Sumer, B.~M. \& Bakioglu, M.} 1984 {On the formation of ripples on an
  erodible bed}. {\em J. Fluid Mech.\/} {\bf 144}, 177--190.

\bibitem[Uhlmann(2005)]{Uhlmann2005a}
{\sc Uhlmann, M.} 2005 {An immersed boundary method with direct forcing for the
  simulation of particulate flows}. {\em J. Comput. Phys.\/} {\bf 209}~(2),
  448--476.

\bibitem[Uhlmann(2008)]{Uhlmann2008}
{\sc Uhlmann, M.} 2008 {Interface-resolved direct numerical simulation of
  vertical particulate channel flow in the turbulent regime}. {\em Phys.
  Fluids\/} {\bf 20}~(5), 053305.

\bibitem[Uhlmann \& Du\v{s}ek(2014)]{uhlmann:13a-nodoi}
{\sc Uhlmann, M. \& Du\v{s}ek, J.} 2014 The motion of a single heavy sphere in
  ambient fluid: a benchmark for interface-resolved particulate flow
  simulations with significant relative velocities. {\em Int. J. Multiphase
  Flow\/} {\bf 59}, 221--243.

\bibitem[Wong \& Parker(2006)]{wong:06}
{\sc Wong, M. \& Parker, G.} 2006 Reanalysis and correction of bed-load
  relation of {M}eyer-{P}eter and {M}\"uller using their own database. {\em J.
  Hydr. Eng.\/} {\bf 132}~(11), 1159--1168.

\end{thebibliography}

\end{document}